\documentclass[a4paper, 11pt]{article}
\usepackage[utf8]{inputenc}
\usepackage{graphicx}
\usepackage{amsmath}
\usepackage{amsfonts}
\usepackage{amssymb}  
\usepackage{mathtools}    
\usepackage{bm}
\usepackage{epsfig}
\usepackage{caption}
\usepackage{cases}
\usepackage[numbers]{natbib}
\usepackage[a4paper]{geometry}
\usepackage{geometry}
\begin{document}

\title{On the diversity of stationary cosmologies in the first half of the twentieth century}

\author{E-A. Dubois  A. F\"uzfa\\
Namur Institute for Complex systems (naXys) \\ Espace philosophique de Namur (esphin)\\ University of Namur, Belgium }
\date{1 january 2019}
\maketitle
\abstract{
Before the establishment of the hot Big Bang scenario as the modern paradigm of cosmology, it faced several early competitors : the so-called stationary cosmological models. There were truly plural and independent approaches
incorporating cosmic expansion but without time evolution of the total cosmological density, thanks to the inclusion of some processes of continuous matter creation. 
We distinguish here three different independent motivations leading to a stationary vision of the universe. 
First, Einstein's concerns on the asymptotic behaviour of gravitation led
him to consider continuous matter creation in a recently discovered unpublished work dated of 1931. 
Second, there is the quest by Dirac and Jordan for a scientific explanation of numerical coincidences in the values of fundamental constants, leading both to a time variation of Newton's constant and spontaneous matter creation. The third one appears in the steady-state theory of Bondi and Gold as the postulate of the Perfect Cosmological Principle according to which the properties of the universe do not depend in any way of the location and the epoch of the observer.  Hoyle developed a mathematical model of spontaneous matter creation from a modification of general relativity that should be considered as a direct legacy of Einstein's and Dirac's approaches.  Hoyle's model allows obtaining a "\textit{wide cosmological principle}" as an end-product, echoing Bondi and Gold's Perfect Cosmological Principle.
Somewhat ironically, many modern key questions in hot Big Bang cosmology, like dark energy and inflation, can therefore be directly related to the physical motivations of the early stationary cosmologies.

\textbf{steady-state cosmology \and large number coincidence \and cosmological principle \and history of cosmology}
}  

\section{Introduction}
\label{intro}

\indent The current cosmological paradigm consists of an evolving universe: it undergoes global expansion from a primordial, infinitely hot and dense state that Hoyle coined later the "\textit{Big Bang}" \cite{Hoyle51}. \\

In an expanding universe, the density of matter must decrease by virtue of conservation laws. This is the stumbling block on which the so-called stationary cosmological models built and stood against the "\textit{Big Bang}" approach. The term "\textit{stationary cosmology}" was first coined by F. Hoyle in his famous paper \cite{Hoyle48}. However, this term has a completely different meaning than what is nowadays referred to stationary space-time. Indeed, the last refers to a manifold with at least one time-like Killing vector, or equivalently that the corresponding metric is not diagonal (in particular $g_{0i}\ne 0\; i=1,2,3; $) and does not explicitly depends on the time coordinate. While the first, the "\textit{stationary cosmologies}" refers to cosmological models based on the basic assumption that some physical properties of the universe are constants over space and time. In this paper, we choose to follow this convention for historical reasons and we warn the reader about the possible confusion between stationary cosmologies and stationary space-times.\\

The first relativistic model of the cosmos is due to Einstein \cite{Einstein17} and de Sitter built his own model soon afterwards \cite{deSitter1617III}.  Einstein's solution took into account the matter content of the universe but added the so-called cosmological constant to the field equations to allow a static space-time. The equilibrium in the model is guaranteed by an exact balance between the radius and the mass of the universe. de Sitter considered a vanishing density of matter in his cosmological model. Despite this apparently physical non-sense, de Sitter's model became, a few years later, the source of considerable interest because it offered a justification for the observations of the redshifts of the spiral nebulae, through the motion of test particles in de Sitter's space-time. 
Although cosmic expansion was not manifest in the original writing of de Sitter's metric, Lema\^{i}tre showed that under an appropriate choice of coordinates, it really was a dynamical cosmological model \cite{Lemaitre27}. \\ 

In 1922, Friedmann \cite{Friedmann22} derived the first dynamical cosmological solutions.
In his 1927 paper \cite{Lemaitre27}, Lema\^{\i}tre found a similar solution that constitutes a compromise between the Einstein's and de Sitter's models and matched it to yet unpublished data presented by Hubble in a seminar that Lema\^{\i}tre has attended, the later even made a numerical estimate of the cosmic expansion rate, which he dubbed, later on \cite{Lemaitre31}, "\textit{Hubble's constant}". This estimate, unfortunately based on poor data, lead to a critical underestimation of the age of the cosmos, of about 2 billion years, that  Lema\^{\i}tre overcame later on by relying on a positive cosmological constant  and an accelerated cosmological expansion in \cite{Lemaitre33}. Overall, Lema\^{\i}tre's cosmological picture of the famous "\textit{primeval atom}" \cite{Lemaitre31Nature} stood on a rigorous mathematical model of the expansion of the universe, which started with some primitive condensed state in which quantum laws must have been at play.\\

Since that time, strong empirical support has emerged in favour of an expanding universe. 
The first was the observation by Hubble of a linear relation between the redshift and the distance of nearby galaxies \cite{Hubble29}.
These measurements are commonly interpreted as the manifestation of a dynamical space-time, and more precisely of an expansion of space.
As we shall see, this interpretation is also shared by all the stationary cosmologies. Indeed, these models do not deny the dynamics of the universe but yet they are stationary because the matter density remains unchanged on cosmological scales. 
\\

Later, Gamow with his student Alpher and other colleagues extended the primitive state envisioned by Lema\^{\i}tre in a series of papers in which they established the link between the expansion of the universe and the formation of atoms and nuclei. They first predicted the existence of a relic electromagnetic radiation from the formation of atoms
and then explained the relative abundance of chemical elements observed in the cosmos through the process of primordial nucleosynthesis (\cite{Gamow46}, \cite{abg}, \cite{Gamow48} and \citep{AlpherHerman48}). 
In 1965, Penzias and Wilson measured a signal \cite{PenziasWilson65} which can be interpreted as the cosmological microwave background predicted by Gamow and his collaborators. Although Lema\^{\i}tre envisioned the existence of cosmic relics in 1933 \cite{Lemaitre33}, he thought they were instead related to cosmic rays.\\

Nowadays, the hot Big Bang scenario of the expanding universe serves as the current  paradigm for cosmic history. If it agrees fairly well with many different observations, it must be noticed that these experimental evidences were found several decades after the beginning of modern cosmology as summarized above. History is too often written by the victors and so historical narratives often tend to overlook the unsuccessful competitors: this is the case for stationary cosmologies. Let us note that is is important, of course, to distinguish the context of discovery and the context of logical justification that are not the same.
\\

The aim of this paper is to distinguish three different paths toward stationary cosmology in  the first half of the twentieth century, each path having its own physical motivation, which we detail in the next section. 
The first one is due Einstein himself \cite{Einstein31} in an unpublished draft, as was recently discovered by O'Raifeartaigh \cite{ORaifeartaigh14AE31}. Following his investigation of the asymptotic behaviour of gravitation, started in 1917 with his introduction of the cosmological constant,  Einstein arrived to the original idea of continuous matter creation. However, Einstein manifestly gave up the idea, but for unmentioned reasons. O'Raifeartaigh  et al.\cite{ORaifeartaigh14AE31} suggest that, when Einstein realised that a successful steady-state model necessitated a change to the fields equations, he abandoned the model as too contrived.
The second path was opened by Eddington \cite{Eddington3031} in his attempt for a mathematical explanation of the value of the fine-structure constant which was then somewhat pursued independently by Dirac and Jordan in their works on large numbers coincidences. In particular, we shall show that Jordan also introduced a stationary cosmological model involving spontaneous matter creation.
The third path we present is constituted by the work of Bondi and Gold in 1948 \cite{BondiGold48} which is often wrongly assimilated to the independent work of Hoyle \cite{Hoyle48} into the "\textit{Hoyle-Bondi-Gold steady state theory}". Indeed, the starting assumption of Bondi and Gold is the Perfect Cosmological Principle according to which the properties of the Universe do not depend neither on the location of the observer nor on the epoch. On his side, Hoyle developed his own mathematical framework as an extension of general relativity implementing spontaneous matter creation to support his stationary vision of the cosmos \cite{Hoyle48} (see also \cite{Kragh1999}). 
Hoyle's work should therefore be considered as a direct legacy of earlier approaches by Einstein, Dirac and Jordan.\\

In the conclusion, we will give a synthetic representation of these different motivations of these early stationary cosmologies. We will also emphasise some of their echoes in modern research, which are now developed in the framework of the old competitor of stationary cosmology: the hot Big Bang scenario.

\section{Two premises of steady-state cosmology}
\label{sec:2}
In preamble, it must be underlined the difference existing between a static and a stationary cosmological model. The first one has no dynamical process. A static universe remains always exactly the same, identical to itself. The second does not deny a possible dynamical process  but remains similar to itself. A stationary universe evolves with time but some observational measures stay constant.\\

The nomenclature "steady-state" or "stationary" is not used to indicate that these models deny the dynamics of the universe predicted by \cite{Lemaitre27} and measured by \cite{Hubble29}; but, it is a way to dub these cosmological models which postulate an unchanging though expanding universe. Indeed, in these steady-state models cosmic expansion is achieved with a constant expansion rate, making Hubble's parameter a fundamental physical constant. 
Evolution is, in some sense, still preserved in the dynamics of space-time and somewhat also on the matter content that is continuously created so the matter density stays unchanged.
\subsection{Einstein's lost paper}
\label{subsec:21}
In the beginning of the twentieth century, Albert Einstein's theories of special and general  relativity allowed the advent of physical cosmology, based on a rigorous background and equations linking together
space, time, matter and energy. Immediately after having published his relativistic equations of the gravitational field \cite{Einstein16}, Einstein applied, in 1917, his new theory to the universe as a whole as he was questioning the asymptotic behaviour of gravitation and the associated boundary conditions \cite{Einstein17}.

Einstein elegantly eluded the problem of choosing appropriate boundary conditions for space-time by considering the geometry of the 3-sphere which does not require specifying asymptotic behaviour nor boundary conditions. 
However, to ensure the hydrostatic equilibrium of the cosmos, Einstein had to extend his field equation through the introduction of a new cosmological constant term to the field equations, then denoted $\lambda g_{\mu \nu}$, where $g_{\mu\nu}$ is the metric tensor encoding the geometry of space-time and 
 $\lambda$ is the cosmological constant (nowadays denoted by $\Lambda$), which can be interpreted as an intrinsic scalar curvature of space-time (i.e. the scalar curvature in vacuum). 
Following Einstein:  "\textit{that term is necessary only for the purpose of making possible a quasi-static distribution of matter, as required by the fact of the small velocities of the stars}"\cite[p.432 of the translation]{Einstein17}.
This new cosmological constant term introduced a new long-ranged force into relativistic gravitation while being compatible with the fundamentals of general relativity: the equivalence principle and general covariance.
The cosmological constant term $\lambda g_{\mu \nu}$ later turned out as a non trivial adding to general relativity, glimpsing at interplay between quantum mechanics and gravitation. However,  Einstein mention the possibility of doing without it, especially in his letter to Weyl \cite{Einstein23} and discarded it in 1931 \cite{Einstein31B}.\\

In 2014, O'Raifeartaigh et al. published some papers \cite[and references therein]{ORaifeartaigh14AE31} concerning a recently discovered attempt by Einstein on a steady state model of universe, a work left unfinished and abandoned before publication \cite{Einstein31}. With the help of some experts, the unknown manuscript was dated as likely written in early 1931. The content of this draft has been studied in several papers, including \cite{Nussbaumer2014}.\\

The following discussion is based directly from the original work of Einstein in 1931, and not on later translation.
Let us start with Einstein's equations 
\begin{equation}
\left( R_{ik}-\dfrac{1}{2}g_{ik}R \right) -\lambda g_{ik}=\kappa T_{ik}
\label{systEinstein}
\end{equation}
where $g_{ik}$ is the metric tensor, $R_{ik}$ the Ricci tensor, $T_{ik}$ the stress-energy tensor, $R$ the scalar curvature, $\kappa=8\pi G/c^4$ and
$\lambda$ is Einstein's notation for the cosmological constant.\\
The so-called Einstein universe, first derived in \cite{Einstein17}, corresponds to a static space-time with identical 3-sphere as spatial slices, in which matter has a constant density $\rho$ in time and space. 
Unfortunately, this solution is well known to be unstable, since the balance is achieved through a fine-tuning of the value of the cosmological constant $\lambda$ to compensate for the homogeneous density $\rho$. Since the former must rigorously be constant to be compatible with the fundamentals of general relativity while the later is not rigorously constant in the cosmos, the equilibrium is unstable under density perturbations\footnote{This instability can be deduced from the classification of dynamical cosmologies by A. Friedmann in 1922 where the author has shown that Einstein's universe sits at an unstable equilibrium point (see Fig. 125-1 of\cite{Friedmann22}).On this subject, \cite{Eddington30} could also be quoted.}.
Although Hubble's results showed a Doppler effect proportional with the distance, which were interpreted by Lema\^{i}tre as inherited from cosmic expansion, a dynamical property of space-time, Einstein described such a model in his 1931 draft as unacceptable, without detailing his reasons.\\

Einstein suggested instead a different model, compatible with Hubble's observations but with a mean density of matter that was constant over time. The mathematics and notations below are those of Einstein's draft \cite{Einstein31} that we chose to reproduce identically here in order to follow Einstein's steps to a stationary cosmology.
Using the metric\footnote{In Einstein's draft, a crossing-out can be seen, he corrected only one on the two expressions of the metric to have a signature (+, -, -, -). Furthermore, it is interesting to notice that Einstein uses covariant coordinates and not the expected contravariant ones.}
\begin{equation}
ds^2=c^2dt^ 2-e^{\alpha t}\left(dx_1^ 2+dx_2^2+dx_3^2\right) 
\label{metric1}
\end{equation}
and modelling cosmic matter through the stress-energy tensor of pressureless perfect fluid
\begin{equation*}
T^{\mu \nu}=\rho u^{\mu}u^ {\nu}
\end{equation*} 
where $u^\mu$ is the fluid 4-velocity which takes the form, once expressed in the fluid comoving frame, $u^1=u^2=u^3=0$ and $u^4=\dfrac{1}{c}\cdot$ Einstein first noted the following, incorrect, field equations (temporal then spatial): 
\begin{eqnarray}
\frac{3}{4}\alpha ^2-\lambda c^2&=& \kappa \rho c^2,\label{AEmistake_b}\\
\frac{9}{4}\alpha^2+\lambda c^2&=&0\label{AEmistake_a}
\end{eqnarray}
where the first\footnote{We also reproduce the mistake in the physical dimensions of the right hand side of Eq.(\ref{AEmistake_b}).} and second equations of the above set correspond to the spatial and temporal components of Eqs.(\ref{systEinstein}), respectively. 
The system Eqs.(\ref{AEmistake_a}-\ref{AEmistake_b}) imposes $\alpha^2=\dfrac{\kappa c^2}{3}\rho$, leading to a constant density. Einstein then argues that, considering a finite physical volume, particles are constantly escaping because of the cosmic dynamics and therefore there should exist some process of continuous creation of matter in order to keep the total matter density constant. Einstein even proposed that this new particles should emerge from space as this last is not empty due to the presence of the cosmological constant $\lambda$. Einstein therefore associated the cosmological constant term $\lambda$ with the creation of matter as, we will see, Hoyle did in \cite{Hoyle48}.\\

Unfortunately, the first equation (\ref{AEmistake_a}) was wrong and Einstein corrected himself and found \footnote{The mistake probably lies in the computation of the Ricci spatial component $R_{ii}=\frac{-3 \alpha^{2}e^{\alpha t}}{4c^2}$ and consequently of the curvature scalar $R=\frac{3\alpha^2}{c^2}$.}
\begin{equation}
-\frac{3}{4}\alpha^2+\lambda c^2=0,\label{AEcorrect}
\end{equation}
which, together with Eq.(\ref{AEmistake_b}) implies that the matter density is indeed constant but trivially zero. The solution of the problem posed by Einstein in this draft  was in fact de Sitter's vacuum solution\footnote{
If we write down the metric Eq.(\ref{metric1}) as
$$
ds^2=c^2dt^2-e^{2\alpha ct}\left(dr^2+r^2d\Omega^2\right)
$$
with $d\Omega^2=r^2(d\theta^2+\sin^2(\theta)d\varphi^2)$ the solid angle element,
one can obtain the so-called Schwarzschild-de Sitter metric
$$
ds^2=\left(1-\alpha^2 \bar{r}^2\right)c^2 d\bar{t}^2-\left(1-\alpha^2 \bar{r}^2\right)^{-1}d\bar{r}^2-\bar{r}^2(d\theta^2+\sin^2(\theta)d\varphi^2)
$$
through the following change of coordinates
\begin{eqnarray}
r&=&\bar{r}\left(1-\alpha^2 \bar{r}^2\right)^{-1/2}e^{-\alpha c\bar{t}}\nonumber\\
ct&=&c\bar{t}+\frac{1}{2\alpha} \log\left(1-\alpha^2\bar{r}^2\right)
\end{eqnarray}
the angular coordinates being unchanged. It could be noticed that in 1917 de Sitter did not already work in the Schwarzschild-de Sitter coordinates \cite{deSitter17}}, once expressed in the coordinates found by Lema\^{i}tre in \cite[and references therein]{Lemaitre25}.\\

Before O'Raifeartaigh's work, it was not known that Einstein himself had considered the possibility of a universe fitted with a continuous process of matter creation, a central feature of the future steady-state models. Einstein's draft should therefore be considered as the first attempt to build a stationary cosmological model within general relativity, through Einstein's proposal to preserve a constant matter density. 
Shortly afterwards, Einstein published a paper on the cosmological problem of the general theory of relativity in which he finally discards both of his 1917 cosmological contributions: (i) his static universe, on the grounds that it is unstable and (ii) the cosmological constant term $\lambda$ since, according to him, Hubble's observations could be accounted for more naturally in general relativity without invoking this term.
This last paper \cite{Einstein31B}, sometimes known as the Friedmann-Einstein model, constitutes the first occasion on which Albert Einstein formally abandoned the cosmological constant term, never to re-instate it afterwards. An analysis and first English translation of the paper has recently been provided by O'Raifeartaigh and McCann \cite{ORaifeartaigh2014}.
Another contemporary paper by Einstein is his famous work with de Sitter \cite{EinsteindeStitter32}, in which they estimated the pressureless matter density of the expanding universe and discussed other measurable quantities in cosmology.
However, there seems to us that there are no connections to make between this published paper \cite{EinsteindeStitter32} and the unpublished one \cite{Einstein31}.

\subsection{Fundamental Constants, Large Numbers Hypothesis and the Stationary Universe}
\label{subsec:22}
Several reflections on the fundamental constants of physics also led some physicists to envision a stationary view of the universe.\\

On the edge of the nineteenth century, in \cite{Planck1899}, Planck build  his famous set of absolute, natural, physical units, by combinations of $c$ the speed of light, $h$ Planck's constant, $G$ Newton's gravitational constant, $e$ and $m_e$ the charge and the mass of the electron respectively. For example, Planck defined a unit of time as $t_P=\sqrt{\dfrac{G\hbar}{c^5}}\simeq 10^{-43}s$ while Dirac and other authors worked with a unit of atomic time\footnote{The unit of measure of the electron charge $e$ in the cgs system was the statcoulomb which has units $\left[ cm^{\frac{3}{2}}g^{\frac{1}{2}}s^ {-1}\right]$.}: $\displaystyle{\frac{e^2}{m_ec^3}}$ .
These absolute units, build on fundamental physical constants, can then be combined to construct some dimensionless numbers.\\ 

In \textit{Relativity theory of Protons and Electrons} \cite{Eddington36}, Eddington worked on "\textit{a series of investigations in the borderland between relativity theory and quantum theory}. (p.V) ". He noticed a multiple appearance of the natural number\footnote{Eddington first worked on the number $136$, later he rather focused on $137$, this change is due to the improvement of the measure of $e$ and $m_e$ } $136$, and in his opinion "\textit{there are no arbitrary constants left in the scale of relation of natural phenomena}." (\textit{ibid} p.V). Eddington has set a theory for calculating each of them purely deductively, especially from his theory of the Clifford algebra.
He thereafter set a line of research of other authors that will search for explanation of the precise values of these constants.\\

Dirac, in \cite{Dirac37}, noticed that the age of the universe, estimated at that time around $2\times10^9$ years, once expressed in atomic units, gives $10^{39}$, and dubbed
this large number the \textit{epoch} $t$.
Furthermore, the ratio $\gamma$ of the strength of the electrostatic (Coulombian) force over the one of the gravitational (Newtonian) attraction between the electron and the proton in the fundamental system of the hydrogen atom has also a value around $10^{39}$. These large numbers, $\gamma$ and $t$, appear in two a priori independent physical contexts so that they should rely on two fundamentally different justifications.
However, such a coincidence that the ratio $\gamma$ is roughly equal to the epoch $t$ might reveal a deep connexion in nature between cosmology and the atomic theory. Following Dirac, this closeness of two independent and such large numbers cannot be purely coincidental, so $\gamma$ must change as the universe evolves, if this coincidence reflects a deeper yet unknown physical law that is forever valid. 
In addition, the other large dimensionless numbers occurring in nature can all be expressed as simple powers of the -non dimensional- epoch $t$. For example, the ratio between the mass of the observable universe and the one of a proton, which is equal to $10^{ 78}$, could be seen as $t ^2$.
As a consequence, Dirac noted that the number of protons and neutrons have to increase as $t^2$ and the Newtonian gravitational constant has to be inversely proportional to $t$.\\

In 1938, Dirac suggested a new starting point for cosmology based on the following principle, that will be later on called Dirac's principle : 
"\textit{Any two of the very large dimensionless numbers in nature can be connected by very simple mathematical relation}"\cite[p.201]{Dirac38}.
He then suggested a new view of the cosmos assuming an infinite space filled with an infinite amount of matter.
If Dirac mentioned the possibility of spontaneous creation or annihilation of matter, he soon rejected it because it does not fit with any idea in theoretical physics: "\textit{There is no experimental justification for this assumption since a spontaneous creation or annihilation of protons and neutrons sufficiently large to alter appreciably the law (3)} [$a(t) \propto t^{1/3}$] \textit{would still be much too small to be detected in the laboratory. However, such a spontaneous creation or annihilation of matter is so difficult to fit in with our present theoretical ideas in physics as not to be worth considering, unless a definite need for it should appear, which has not happened so far, since we can build up a quite consistent theory of cosmology without it}"(\textit{ibid} p.204)
Dirac therefore abandoned for a while this model because of the success of the standard paradigm, before he
went back to it in the seventies, \cite{Dirac73},\cite{Diracvatican} and \cite{Dirac74}.\\

However, while Dirac gave up this stationary cosmology, someone else soon took it over: 
 P. Jordan developed a completely new cosmological model in a series of papers \cite{Jordan37}, \cite{Jordan38}, \cite{Jordan39}, \cite{Jordan48}.
His motivation was simple: it is useful for science to explore every possible solution, in particular for what concerns the cosmological problem, and not stay confined to what currently appears as good explanations.\\

Jordan, in \cite{Jordan37}, followed the way opened by Dirac and Eddington. He made the calculation\footnote{Jordan assumed a universe ten times older than Dirac's one, compatible with the age of the Earth known by geological data at that time.
The difference in their calculations must reside in the precision of $e$ and $m$.} 
to express the epoch in atomic unity, arriving at the huge value of $10^{41}$.
He suggested that we can neglect the daily growth of the epoch, because this represents only a modest increase of $10^{27}$ time atomic units per day. Jordan saw in his study of the large dimensionless constants a possibility to reconcile astronomical and atomic physics. 
He pursued this idea in several papers, such as \cite{Jordan38}, \cite{Jordan39}, \cite{Jordan48}.
Jordan based his model (i) on the principle previously introduced by Dirac that the dimensionless numbers in nature must be mathematically related and (ii) on the unity and coherence of the universe (while still incorporating cosmic expansion).\\

This cosmology requires a creation of some matter content, which Jordan considered to be stars, a small amount of matter compared to the vastness of matter in the universe. 
However, Jordan's explanation of spontaneous creation of \textit{simple stars} is based on energy arguments and remains quite vague: the net energy cost of the creation of a star is null because the negative gravitational binding energy is balanced by the rest energy. "\textit{In an Euclidian free-mass space, the spontaneous creation of a spherical mass $M_0$ of constant density and with a radius $R_0$ requires none energy, if $R_0=\frac{3}{40\pi} \kappa M_0 $. Because, to scatter this sphere entirely against the gravitation, the same energy $M_0c^2$ would be necessary, that could be represent by these scattered masses.}"\cite[p.69]{Jordan39}\footnote{This citation is extracted from our translation of Jordan's works from the German.} \\
 
In 1949, Max Born drew attention of the scientific community to Jordan's works via a publication in Nature \cite{Jordan49}. 
Paul Couderc in his work \cite{Couderc50}, studied Jordan in the appendix and classified it as a heterodox theory.
Indeed, it must be recognized that Jordan's approach was quite heuristic and not supported rigorously by some mathematical background, at the opposite of Hoyle's approach in 1948.

\section{The Twofold Steady-State Cosmology}
\label{sec:3}
 One cannot study the stationary cosmologies without considering two different approaches that are nowadays referenced collectively as the Bondi-Gold-Hoyle steady-state model.
H. Bondi, T. Gold and F. Hoyle spent a lot of time together during the WWII and, according to \cite{Mitton2011} and \cite{Livio2013}, the idea of an evolving but unchanging universe pried into their mind after their vision of the movie \textit{Dead of Night} \cite{film45}.
This lead these authors to two separate publications.
Indeed, in the fifth release of 1948 MNRAS, you can find these two papers: \textit{The Steady-State theory of the expanding universe} by H. Bondi and T. Gold \cite{BondiGold48}, submitted in July, followed by \textit{A New model for the expanding universe} by F. Hoyle \cite{Hoyle48}, submitted in August.
These two articles presented crossed references and commentaries; however the two approaches expressed are quite different epistemologically and mathematically.\\

First of all, it must be noticed that Bondi and Gold quoted Hoyle's paper. 
As for him, Hoyle introduced his paper revealing that its general idea came from a discussion with Gold and thanking Bondi for his comments and their many discussions. At first sight, one could wonder why they did not publish together.
The reason is that, beyond their friendship, the ideas defended by the authors are, if not opposite, at least very different, and drove them to a kind of schism. 
In a later discussion of \cite{HoyleNarlikar63mach}, Hoyle is questioned by Schlegel about the Perfect Cosmological Principle, an idea developed in \cite{BondiGold48}; Hoyle's answer is welcomed by this self-explanatory comment by Bondi: "\textit{We do not all agree}" \cite[p.340]{HoyleNarlikar63mach}. 
This disagreement is also underlined by \cite[pp. 179-186]{Kragh1999} and in \cite{Kragh2015}. 
The assimilation of theses two articles in a unique Steady-State Theory proposed by these three authors is a misunderstanding or, at least, a late assimilation and shortcut.

\subsection{The Perfect Cosmological Principle of Bondi and Gold}
\label{subsec:31}
Bondi and Gold suggested a heuristic approach based on the \textit{Perfect Cosmological Principle}: "\textit{As the physical laws can not be assumed to be independent of the structure of the universe, and as conversely the structure of the universe depends on the physical laws, it follows that there may be a stable position.}" \cite[p.254]{BondiGold48}.
The Perfect Cosmological Principle therefore shares the conviction of the universality principle according to which the laws of physics do not change with time and stay everywhere valid. To assume this principle lead Gold and Bondi to a universe that is homogeneous and stationary at large scales.
As the they noted, the opposite hypothesis led to an untenable situation "\textit{We do not claim that this principle must be true, but we say that if it does not hold, one's choice of the variability of the physical laws becomes so wide that cosmology is no longer a science}." \cite[p. 255]{BondiGold48}. With this assumption, one can be fully confident about the universality of observations and the validity of their interpretations.\\

Considering a homogeneous and stable universe, one cannot reject the observed large-scale motions, so that the model must incorporate cosmic expansion.
"\textit{It is clear that an expanding universe can only be stationary if matter is continuously created within it.}"\cite[p. 255]{BondiGold48}.
Such an expanding universe can be stationary only if you allow continuous creation of matter.
Bondi and Gold then mentioned that such a model must be described by a de Sitter metric, as Einstein did in his draft in 1931 and as Hoyle did in his 1948 paper. Indeed, de Sitter's geometry written in the coordinates of Eq.(\ref{metric1}) 
incorporates cosmic expansion at a constant expansion rate $\alpha$ but requires a constant total density.
Since matter density decreases due to cosmic expansion, this must be compensated by continuous creation of matter. As mentionned by Einstein and de Sitter in 1932 \cite{EinsteindeStitter32}, only matter density and cosmic expansion rate were considered as observables at that time, therefore de Sitter's geometry was unavoidable to implement the perfect cosmological principle.

Bondi and Gold then suggested some observational tests, even if they claim the creation process is too faint to be measured. They gave an estimation of the rate of matter creation of about $10^{-43}g$ per second per cm$^3$, corresponding to one new atom of hydrogen per cubic metre per $3\times 10^5$ years. \cite[p.266]{BondiGold48}.
Bondi and Gold also considered the physical process of creation, but in a very qualitative way without a mathematical formulation. Their model was not formulated within the context of general relativity, because Bondi and Gold questioned the assumption that general relativity was valid on cosmic scales. However, they noted that the process of matter creation implies a privileged direction along the time dimension as described in Weyl's causal postulate. Hoyle will implement this mathematically in his theory.
The authors were confident in the possibility of finding a mathematical formulation for their model in the context of field theory but they said "\textit{we have no hesitation in rejecting Hoyle's theory, although it is the first and at the moment only field theory formulation of the hypothesis of continuous creation of matter.}" \cite[p.269]{BondiGold48}. This rejection is especially due to the insertion by Hoyle of non-uniformities.\\

It could be noticed how confusing is this formulation of the perfect cosmological principle. It seems that Bondi and Gold do not make the distinction between the physical laws and there solutions. It is now completely accepted that if equations are symmetric, the solution, resolving of a break of symmetry, cannot be. Bondi and Gold developed an essay without a formal mathematical model and it is something that could be desired by modern readers.

\subsection{Hoyle's mathematical approach to stationary cosmology}
\label{subsec:32}
 Hoyle's model \cite{Hoyle48} was based on the line of thought initiated by Jeans \cite{Jeans28} and Dirac \cite{Dirac37} and  developed the idea of a continuous creation of matter for which Hoyle offered a mathematical formulation.
After a study of Newtonian universes, he then turned to the framework of general relativity and, after some general considerations, worked with the following line element\footnote{In what follows, we have corrected some notations used by Hoyle.}:
\begin{equation}
ds^2=c^2dt^2-a^2(t)dl^2,
\end{equation} 
where $dl^2=(dx^1)^2+(dx^2)^2+(dx^3)^2$ is the elementary length of flat Euclidean space.
Following the standard procedure, Hoyle wrote down the non-vanishing Christoffel symbols and components of the Ricci tensor, as well as the scalar curvature. Then he chose to "\textit{diverge from the usual procedure}" to introduce a new mathematical term in Einstein's equation. This term will describe the process of continuous matter creation.\\

Indeed, Hoyle added a space-time four-vector field $C_{\mu}$ whose norm is fixed along time-like geodesics:
\begin{equation}
 C_{\mu}=3\aleph(1,0,0,0),
 \label{vec_C}
 \end{equation}
where $\aleph$ is a constant.
This vector $C_{\mu}$ has vanishing spatial components for symmetry reasons and compatibility with the chosen metric. 
Then, Hoyle constructed a symmetrical tensor $C_{\mu \nu}$ by covariant differentiation of the four-vector field, 
\begin{equation}
C_{\mu \nu}=\triangledown_{\nu}C_{\mu}\cdot
\end{equation}
Given the above-mentioned symmetries, this "\textit{creation}" tensor $C_{\mu \nu}$ counts only three non-zero components:
\begin{equation}
C_{ii}=-3\dfrac{\aleph}{c}a\dot{a}\; ; \; i=1,2,3 \cdot
\end{equation}
Hoyle then introduced the following \textit{modification} of the Einstein field equations: 
\begin{equation}
R_{\mu \nu}-\dfrac{1}{2}g_{\mu \nu}R+C_{\mu \nu}=-\kappa T_{\mu \nu}\cdot
\label{systHoyle}
\end{equation}
Doing this, the stress-energy tensor $T_{\mu \nu}$ is not conserved anymore. Indeed, from the second Bianchi identity, we have:
$$
\nabla_\mu \left(R^{\mu \nu}-\dfrac{1}{2}g^{\mu \nu}R\right)=0.
$$
Then we find that the creation tensor is the source of the production of new particles:
\begin{equation}
\nabla_\mu T^{\mu \nu}=-\frac{1}{\kappa}\nabla_\mu C^{\mu \nu}=-\frac{1}{\kappa}g^{\alpha\nu}\nabla_\mu\nabla_\alpha C^\mu\cdot
\end{equation}
One can see that, if $\aleph =0$, we find back an Einstein-de Sitter universe for a pressureless matter universe with varying $\rho$, while, here, $\rho$ is constant. As $\nabla_{\mu}T^{\mu \nu}\neq 0$, the stress-energy tensor is not conserved and the matter geodesics will be modified, $C_{\mu}$ acting like a long-ranged force. Therefore, in Hoyle's theory, one expects the free fall of matter particles to be affected by this new long-ranged force on cosmological distances. Exploring the observable constraints of this phenomenological property would be a significant and interesting work  which goes beyond the scope of the present paper. \\

Comparing the equations (\ref{systEinstein}) and (\ref{systHoyle}), one can notice a similarity between the cosmological constant term and the creation tensor. 
In 1951, Mc Crea \cite{McCrea51} included the creation tensor $C_{\mu \nu}$ directly into the total stress-energy tensor on the right-hand side of Einstein equations so that their form stayed unchanged. Thereby, he rightly claimed there is no net creation of matter since the total stress-energy tensor is conserved, due to the second Bianchi identity. However, this yields some coupling between usual matter and a source of usual matter, given by $C_{\mu \nu}$, and so there is indeed no conservation of the number of usual matter particles.\\

It is important to notice that Hoyle put the term $C^{\mu \nu}$ responsible of the creation of matter on the left hand side of his modified Einstein equations, as a modification of the geometry in the field equations. This approach is similar to Einstein's one in 1917, with his introduction of the cosmological constant term, with the important difference that the modification brought by Hoyle does have an impact on matter conservation while this is not the case for the cosmological constant term. Matter conservation laws are the same whatever the value of the cosmological constant while matter conservation laws must include the creation term in Hoyle's approach. Yet both approaches by Einstein in 1917 and Hoyle in 1948 consisted of modifying the way space-time is curved by matter to account for cosmological considerations.\\

In addition, Hoyle's model allows matter creation without introducing any violation of the fundamental laws of covariance and the equivalence principle. Gravity is still described through only one metric in the Levi-Civita connection, so that the total stress-energy and Einstein's tensors are conserved. 
Matter creation can be seen as the result of some interaction between usual matter and some invisible sector through the term $C_{\mu\nu}$. In \cite{Hoyle48}, this sector is modeled in cosmology by some frozen vector field Eq.(\ref{vec_C}).\\

In the metric chosen by Hoyle, the modified Einstein field equations become (temporal equation first, then the spatial one):
\begin{eqnarray}
3\dot{a}^2 = \kappa \rho c^4a^2\cdot && \\
2a\ddot{a}+\dot{a}^2-3\aleph ca\dot{a}&=&0
\end{eqnarray}
It is possible to show that this set of equations has a first integral that takes the form of a Bernouilli equation. The general solution is 
\begin{equation}
a(t)=\left(C_1+C_2e^{\frac{3}{2}\aleph ct}\right)^{\frac{2}{3}}
\label{solconvode}
\end{equation}

The particular solution given by Hoyle, supposing $\dfrac{\dot{a}}{a}=\aleph c$ at $t=0$, is
\begin{equation}
\begin{cases}
\frac{3}{4}\aleph ct=\tanh^{-1} \left( \frac{2\dot{a}}{a\aleph c}-1 \right) -\tanh^{-1} \left(2\alpha-1 \right) \text{, if }\alpha <1,\\
\frac{3}{4}\aleph ct=\coth^{-1} \left(\frac{2\dot{a}}{a\aleph c}-1\right) -\coth^{-1} \left(2\alpha-1 \right) \text{, if }\alpha >1.
\end{cases}
\label{solFH}
\end{equation}
Isolating $\frac{\dot{a}}{a}=d(\log(a))/dt$ in (\ref{solFH}), it becomes easy to develop these expressions to have the general solution. For the first equation of (\ref{solFH}), $-\dfrac{a}{c}C_1 = \dfrac{1}{2}e^{\frac{3}{2}D-C}$ and $-\dfrac{3c}{a}C_2 = \dfrac{1}{2}e^{\frac{3}{2}D+C}$ where $C=\tanh^{-1}(2\alpha -1)$ and $D$ is an integration constant.\\
 
It is interesting to notice that the general solution (\ref{solconvode}) of Hoyle's model only reduces to de Sitter space-time and a constant Hubble parameter for this particular choice of initial conditions Eq.(\ref{solFH}) (or $C_1=0$ in (\ref{solconvode})) or, equivalently, $a\rightarrow 0$  if $t\rightarrow -\infty$. 
This is the only choice allowing both constant density and Hubble parameter, since if $a\rightarrow a_{-\infty}\neq  0$  when $t\rightarrow -\infty$, one finds
a solution with varying density and Hubble parameter. Hoyle's steady-state cosmological model is therefore a consequence of a modification of gravity and a particular choice of initial conditions to guarantee the constancy of $H_0$ as was considered by many at that time.\\

Hoyle came to this conclusion: "\textit{It is only through the creation of matter that an expanding universe can be consistent with conservation of mass within the observable universe}" \cite]p.379]{Hoyle48}. 
In standard cosmology, based on general relativity, it is the stress-energy tensor that is conserved through the conservation laws $\nabla_\mu T^{\mu \nu}=0$. 
Hoyle instead requires that it is the \textit{mass} that is conserved all along cosmic expansion, which is very different: mass is a global quantity \cite{Blanchet2011} while stress-energy is a local one. 
To ensure this conservation of mass during cosmic expansion, Hoyle has modified Einstein's general relativity but in such a way that there are still conservation laws, but only for the total stress-energy tensor $T_{\mu\nu}+\frac{1}{\kappa}C_{\mu\nu}$. 
In addition, the matter creation process must exactly balance the decrease of matter density due to cosmic expansion, which imposes some symmetries to the vector field $C_\mu$.\\

McCrea \cite{McCrea51} suggested a satisfactory description of the consequences of the creation hypothesis may nevertheless be obtained without any modification of Einstein's equations. 
The creation tensor $C_{\mu \nu}$ is absorbed in the stress-energy tensor.\\

Hoyle's model has an infinite past and an infinite future, it also satisfied the wide (or, in Bondi and Gold's wording, the perfect) cosmological principle. 
However, this is a consequence of his development when it was the starting hypothesis of the Bondi-Gold model.
Hoyle's stationary model emerges from a modification of general relativity that still incorporates conservation laws but for an extended definition of the matter content of the universe. Therefore, Hoyle's approach is also reminiscent of the one made by Einstein in 1917, when the father of general relativity introduced his famous cosmological constant. 

\section{Conclusion}
\label{ccl}
In parallel with the development of the hot Big Bang cosmological model, some other theorists, including Einstein, Dirac, Jordan and later on Bondi, Gold and Hoyle, explored the idea of a stationary universe. Although their motivations were quite different, they all finally came to the need for a process of continuous matter creation.\\

\begin{figure}[h!]
\centering
\includegraphics[scale=0.25]{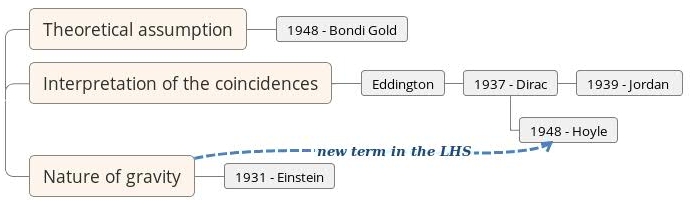}
\caption{Three ways to stationary cosmology}
\label{illu}
\end{figure}

Einstein attempted a steady-state model, apparently in early 1931, by questioning the nature of gravity and exploring the possibilities offered by his equations. He however quickly abandoned the idea when he realized his approach led to a trivial solution with vanishing density and that a successful approach would require an alteration to the field equations. This 1931 approach was similar to the reasoning that led him to introduce the famous cosmological constant term in his 1917 paper. Dirac and Jordan's approaches followed original work of Eddington in 1936 \cite{Eddington36} who tried to explain  the numerical coincidences between atomic physics and astronomy.  
Finally, what is usually referenced as steady-state theory actually comes from two distinct models of Hoyle on the one hand, and of Bondi and Gold on the other. The first model by Hoyle can be seen as a link between the exploration of the nature of gravity by Einstein and Dirac's investigation on large numbers. For the second, the motivation is purely philosophical: Bondi and Gold wanted to save the space and time invariance of the physical laws.
These different motivations behind stationary cosmology are summarized in (Fig. \ref{illu}).

\newpage
In the first half of the twentieth century, the so-called Big Bang model and the steady-state theory emerged as candidates for the status of paradigm of physical cosmology. Even though strong evidence in support of the Big Bang model eventually emerged, notably through the detection of the cosmic microwave background in 1965, the stationary cosmological models, though empirically discarded still are of interest through their underlying epistemological questions and their physical process and mathematical tools that are still in use today. Remarkable examples are the use of de Sitter in inflationary cosmology, and spontaneous creation coupling terms which are currently used in the description of coupled dark energy-dark matter models \cite{Amendola2010}, \cite{AF2014} and \cite{Coppeland2006}.\\

In the last decades, the hot Big Bang scenario has been extended to account for many new observations (galaxy rotation curves, relative height and angular size of the acoustic peaks in the CMB, baryon acoustic oscillations, distance measurements of type Ia supernovae, galaxy clustering, etc.) and to solve the horizon and flatness problems. This leads to the notion of extending matter content by adding new ingredients of dark matter and dark energy to which ordinary matter could directly couple, which would imply its non-conservation during cosmic expansion as with matter creation processes in steady-state cosmology.\\

The later developments of physical cosmology in the XXth century eventually settled the hot Big Bang scenario as the paradigm of describing the Universe at large, while stationary cosmologies were progressively abandoned, yet not completely. There is large literature detailing the scientific reasons behind the emergence of the current paradigm at the detriment of stationary cosmologies, and a good starting point is certainly the books by Kragh \cite{Kragh1999} and Longair \cite{Longair2006} and references therein.\\

Quite interestingly, many key investigations of modern cosmology already appeared in the very first stationary cosmologies, as we presented here. One can actually relate the pioneering works on stationary cosmologies to four important modern research areas that have interconnections: equivalence principle, modified gravity, cosmological principle and the origin of ordinary matter in the universe. 
First, Dirac and Jordan's approaches considered the variation of fundamental physical constants which leads to violation of the equivalence principle. Indeed, since those constants are involved into the binding energies of objects, any such change in the physical constants will result in a variation of inertial or gravitational masses and non-universality of free fall. In addition, a space-time variation of fundamental constants must be described by consistent field theory, such as Jordan-Fierz-Brans-Dicke tensor-scalar gravity (\cite{Jordan49}, \cite{Jordan59}, \cite{Fierz56} and \cite{BransDicke61}). Therefore, the approach of stationary cosmologies by Dirac and Jordan finally lead to question the equivalence principle and to formulate extensions of general relativity. 
Second, Einstein's and Hoyle's starting point was a modification of general relativity that incorporates spontaneous matter creation and from which it is possible to derive a stationary cosmological model as a particular solution. This particular case is actually de Sitter solution in which the cosmological singularity is retrieved asymptotically in the past, as in modern models of eternal inflation\cite{Aguirre2003}.
Third, Bondi and Gold's approach to stationary cosmology is related to the debate on the cosmological principle, although nowadays it is focused today on the feedback of large-scale inhomogeneities on the background cosmic expansion. 
Finally, all the approaches to stationary cosmologies that we described here imply spontaneous matter creation and therefore question the origin of ordinary matter in the universe. This is still an open question at the heart of several research areas including particle creation during inflationary processes, baryo- and leptogenesis, dark matter creation from radiation, etc.(\cite{Allahverdi2010}, \cite{Ringeval2015}, \cite{Ringeval2016} and \cite{Clesse15}).\\

Although these four research areas have been widely active in the last four decades for tackling several cosmological problems of the hot Big Bang paradigm, it is interesting to see how these ideas also emerged originally in attempts to build stationary cosmologies. Today, we know how modifications of gravity, violation of the equivalence principle (\cite[and references therein]{Damour18} and \cite{Planck2016}) and departures from the cosmological principle are well constrained (\cite{Planck2016bis} and \cite{Planck2016ter}).\\
 
However, the processes at work to make the universe stationary, like the creation of matter needed to compensate the decrease of density from cosmic expansion, are also expected to be very faint. It could interesting for future work to examine how Hoyle's modification of gravity for stationary cosmology and the related phenomenological consequences presented here are constrained by modern observations.
It is also interesting to notice that several of the original authors pursued their investigations around stationary cosmology for decades. Dirac came back to the implications of the large number hypothesis on cosmology in 1973 (\cite{Dirac73}, \cite{Diracvatican} and \cite{Diracvatican}). Jordan further developed alternative theories of gravity in five dimensions \cite{Jordan48} and investigate geological and planetological evidence in favour of a variation of Newton's constant \cite{Jordan59}. Hoyle never abandoned stationary cosmology and continue to improve his mathematical model and its phenomenological consequences (\cite{HoyleNarlikar64cfield},\cite{HoyleWickramasinghe67Natb} and \cite{Hoyle93}).\\

The deep origin of stationary cosmologies is clearly not motivated by the quest for cosmological models avoiding Big Bang singularity, rooted in philosophical assumptions against the implications of the existence of such a singularity or in new observations.  At the opposite, stationary cosmologies appeared at a time where the Big Bang scenario was not favoured as it is today and were based on a study of the consequences of fundamental principles framing all physics.
And, despite of the opposition of these two cosmological visions, the evolutionary against the stationary one, they actually share common motivations and applications. The current cosmological paradigm strongly benefits from these early investigations with profound physical motivations, which should therefore not be considered as mere historical curiosities or misguided ways. 

\section*{acknowledgements}
The authors would like to thank D. Bertrand for his precious help in translation of German works, especially by Jordan; we are also grateful to D. Lambert and S. Clesse for useful comments and support.

\bibliographystyle{spmpsci}

\end{document}